# Reconfigurable Spin-Wave Properties in Two-Dimensional Magnonic Crystals Formed of Diamond and Triangular Shaped Nanomagnets

Swapnil Barman[1] and Rajib Kumar Mitra[2,*]

[1]Technical Research Centre, S. N. Bose National Centre for Basic Sciences, JD Block, Sector III, Salt Lake City, Kolkata – 700106, India

[2]Department of Chemical and Biological Sciences, S. N. Bose National Centre for Basic Sciences, JD Block, Sector III, Salt Lake City, Kolkata – 700106, India

Email: rajib@bose.res.in

**Abstract**

Two-dimensional ferromagnetic nanodot structures exhibit intriguing magnetization dynamics and hold promise for future magnonic devices. In this study, we present a comparative experimental investigation into the reconfigurable magnetization dynamics of non-ellipsoidal diamond and triangular-shaped nanodot structures, employing broadband ferromagnetic resonance spectroscopy. Our findings reveal substantial variations in the spin wave (SW) spectra of these structures under different bias field strengths ($H$) and angles ($\varphi$). Notably, the diamond nanodot structure exhibits a variation from nearly symmetric W-shaped dispersion to a skewed dispersion and subsequent transition to a discontinuous dispersion with subtle variation in bias field angle. On the other hand, in the triangular nanodot array a SW mode anti-crossing appears at $\varphi = 15°$ which is starkly modified with the increase in $\varphi$ to $30°$. By analyzing the static magnetic configurations, we unveil the nature of the SW spectra in these two shapes. We reinforce our observations with simulated spatial power and phase maps. This study underscores the critical impact of dot shape and inversion symmetry on SW dynamical response, highlighting the significance of selecting appropriate structures and bias field strength and orientation for required functionalities. The remarkable tunability demonstrated by the magnonic crystals underscores their potential suitability for future magnonic devices.

# 1. Introduction

The functionality of modern electronic devices is heavily reliant upon integrated circuits, which comprise a very high number of transistors embedded in them. As the miniaturization of devices continues, the density of transistors increases exponentially. Various challenges arise due to this, such as the thermal runaway problem[1]. Alongside this, future technology demands higher speeds of devices as well. Research domains such as magnonics[2, 3], spintronics[4, 5] and magnon spintronics[6] have emerged in recent decades, as promising alternatives to charge-based technology. Nanomagnetism[7, 8] has potential applications in magnetic memory[9, 10], data storage[11, 12], and logic devices[13, 14], along with spin torque nano-oscillators[15] and on-chip microwave communication devices constituting magnonic crystals (MCs)[16], which are systematically patterned ferromagnetic nanostructures. These developments have us progressing towards all-magnetic computing. The domain of magnonics utilizes spin waves (SWs) as information carriers across MCs. For the efficient design of magnonic devices, it is important to understand their static and dynamic magnetic properties. The SW dynamics[17] in these MCs can be controlled by tuning various extrinsic and intrinsic parameters. The intrinsic properties that administer the SW dynamics in these ferromagnetic nanostructures include dot size[18, 19], shape[20], lattice symmetry[21], lattice constant[22], and choice of material. Conversely, the external parameter includes the strength as well as orientation of the bias magnetic field which can be varied. Additionally, SW properties can be tuned by spin transfer torque, microwave power and temperature. MCs display various static spin configurations (microstates) and SW properties that are reconfigurable by tweaking the parameters mentioned.

In the realm of magnonics, a pivotal endeavour revolves around achieving a profound comprehension of SWs and mastering their manipulation in patterned arrays of nanomagnets. Remarkable advancements in nanofabrication and high-frequency measurement techniques have assisted in an era of exploring SW dynamics within intricate, patterned magnetic nanostructures. Over the past decade, an abundance of research has delved into SW propagation[23, 24], localization[25, 26], hybridization[27], and damping[28, 29] across an array of patterned structures. These structures include periodically patterned magnetic arrangements such as dot lattices[30, 31], nanowires[32], nanostripes[33], and nanochannels[34] among others. The primary objective in scrutinizing these patterned magnetic nanostructures is to gain insight into, and subsequently fine-tune various aspects of SW properties. This encompasses the SW frequency, the inherent nature of modes, the profiles of these modes, and the associated damping. In recent years, significant attention has been directed toward the study of two-dimensional ferromagnetic nanodot arrays, yielding valuable insights into the modes of SWs and their anisotropic characteristics. Micromagnetic simulations have demonstrated tunable spin-wave dynamics in $Ni_{80}Fe_{20}$ (Permalloy, Py hereafter) nanodot arrays with different dot shapes, revealing quantized modes and inter-dot interactions for varying edge-to-edge dot separations[35]. Py dots have also shown quasi-1D edge spin waves in buckle state and 2D waves in Y state[36], promising magnonic applications. Triangular nanodot arrays demonstrate enhanced coercivity, in-plane anisotropic magnetization, and local magnetic shape anisotropy[37]. Normal modes in equilateral triangular nanodots with vortex, Y, and buckle spin states reveal distinct frequencies, profiles, and nodal surface behaviours through micromagnetic representation, with Brillouin light scattering spectra calculated for magnetic microstates[38]. Additionally, measurements on Py triangular nanomagnets have uncovered diverse SW modes and hybrid configurations[25].

In addition, theoretical studies have delved into the normal modes of SWs in circular ferromagnetic nanodots, revealing a significant influence of the dipole exchange interaction[39]. Cylindrical honeycomb structures exhibit Dirac magnons[40], while two-dimensional magnonic crystals demonstrate nonuniform SW softening, offering a pathway for omnidirectional magnonic band gap creation[41]. The reversible tuning of these band gaps using magnetic fields and in-plane squeezing structures[42] has been explored. Circular and elliptical Py nanodots, studied via the matrix method and Brillouin light scattering, show intriguing mode symmetry and localization[43]. SW excitation through spin-polarized currents in magnetic nanopillars[44] and theoretical investigations into cylindrical dots with vortex states[45] contribute to our understanding of nanoscale spin dynamics. This collective research framework sets the stage for our current exploration of nanomagnetic structures and their potential applications in magnonics. However, an evident gap exists within the existing body of literature. To date, no reported studies have comprehensively addressed the reconfigurability of magnetic field dispersion in the realm of SW dynamics for diamond and triangular nanodots.

Here, we present a comparative study of the role of shape anisotropy and inversion symmetry in Py diamond and triangular-shaped nanodot arrays arranged in square lattices, with the aid of broadband Ferromagnetic Resonance (FMR) spectroscopy. Py is selected due to its negligible magneto-crystalline anisotropy and small coercivity[46]. Upon investigation, we observe large reconfigurability in SW dynamics of diamond and triangular nanodot arrays, by manipulating the azimuthal angle of the in-plane bias field and by cycling the magnetic field from positive to negative direction and vice versa. Further, micromagnetic simulations were conducted to qualitatively reproduce the experimental data. The static spin configurations at various bias fields were used to demonstrate transition phases. In due course, power and phase profiles were numerically calculated to support the observed behaviour. Magnetic hysteresis loops were simulated to understand the magnetic switching behaviour and correlate it with the variation in the SW dynamics. The above study has demonstrated a stark difference in the SW spectra and mode profiles of the two nanodot shapes at different bias field angles due to their shape anisotropy and the ensuing variation of spin textures with the bias magnetic field variation. Particularly the observation of continuous variation of field dispersion of diamond shaped nanodot from a nearly symmetric W-shaped dispersion to a skewed dispersion and subsequent transition to a discontinuous dispersion with subtle variation in bias field angle is remarkable. The magnetic field dispersion is also reversed by reversing the magnetic field cycling direction. These observations will aid in the design of efficient reconfigurable magnonic devices, contributing to the ongoing pursuit of all-magnetic computing and advanced information processing.

1. **Experimental and Simulation Details**
A. **Sample Fabrication**

Two 20-nm-thick Py nanodot arrays, organized in square lattice symmetry, consisting of diamond and triangular-shaped dots respectively, were fabricated on Si(100) substrate via a combination of electron-beam lithography (EBL) and electron-beam evaporation (EBE) techniques. The Py layer is deposited on the Si(100) substrate in a high vacuum chamber at a base pressure of $2 \times 10^{-8}$ Torr. The arrays are (20 μm × 200 μm) in size, with the dots having height and width of 300 nm (±10 nm) and 330 nm (±10 nm). The dots are spaced at an edge-to-edge interval of 75 nm (±10 nm). A coplanar waveguide (CPW), made of Au, with 150 nm thickness, 300 μm length, 25 μm central conducting width (w) and 50 Ω characteristics impedance ($Z_0$) was meshed on top of the diamond as well as triangular-shaped dot arrays, for the broadband FMR measurements. A 5-nm-thick protective layer of Ti was subsequently integrated on top of the Au layer, both of which were done at a base pressure of $6 \times 10^{-7}$ Torr. The CPW was patterned via maskless optical lithography, and is separated from the nanodot arrays by a 60-nm-thick insulating layer of $Al_2O_3$ deposited on top of the Py layer, via EBE. The field emission scanning electron microscope (FESEM) images in Fig. 1(a) and (b) display the diamond and triangular-shaped nanodot arrays respectively and the slight asymmetry in the elements along with rounded corners. These defects have been taken into consideration for the micromagnetic simulations.

**B. Measurement Techniques**

The FMR measurements to study the SW spectra were conducted using a high-frequency broadband vector network analyser (VNA, Agilent PNA-L, model no.: N5230C, frequency range: 10 MHz to 50 GHz), a home-built high-frequency probe station equipped with a non-magnetic ground-signal-ground (G-S-G) picoprobe (GGB Industries, model no.: 40A-GSG-150-EDP) and a coaxial cable. A microwave signal of power -15 dBm and a range of frequencies is applied to the CPW and the reflected signal is transmitted back to the VNA via the G-S-G probe and the co-axial cable. The characteristic SW spectra of the samples are obtained from the frequency-dependent real part of the scattering (S)-parameter (Re($S_{11}$)) in the reflection geometry. The bias magnetic field (H) in the x-y plane was varied in an angular range of $0° \leq \varphi \leq 30°$ with a rotating electromagnet in the range $-1.6 \leq H \leq 1.6$ kOe. The measurements were taken at intervals of 20 Oe and were carried out at room temperature, with a cyclic variation from positive (+ve) to negative (-ve) as well as negative to positive magnetic field. The measured SW spectra for the two nanodot arrays are represented in Fig. (2) and (3).

**C. Micromagnetic Simulations**

Micromagnetic simulations were performed on the nanodot arrays, using finite-difference method based mumax$^3$ software[47], to reproduce the SW modes obtained from the experimental results. The FESEM images of the structures were mimicked, and a two-dimensional (2D) periodic boundary condition (PBC) was applied in consideration of the wide array of elements during the experiment. The arrays were discretized into identical rectangular cells of dimension $4 \times 4 \times 20$ nm$^3$. The cell size was kept below the exchange length of Py (5.2 nm) for the encompassment of exchange interaction. The material parameters used in the simulation were gyromagnetic ratio ($\gamma$) = 17.95 MHz/Oe, saturation magnetization ($M_s$) = 800 emu/cm$^3$, Gilbert damping constant ($\alpha$) = 0.008 (for dynamic studies)[19], exchange stiffness constant ($A_{ex}$) = $1.3 \times 10^{-6}$ erg/cm[19], and magnetic anisotropy field ($H_k$) = 0, for permalloy. The parameters $\gamma$, $M_s$ and $H_k$ were extracted from the Kittel fit of bias field-dependent spectrum of 20-nm-thick Py film as described in the following section of this article, while $A_{ex}$ is taken from the literature[48]. Figs. (2) and (3) represent the reproduced experimental results, along with additionally calculated hysteresis loops. Further, we have numerically calculated spatial maps of the power and phase profiles of the structures using a custom-built MATLAB code,

DOTMAG[49]. Figs. (4) to (7) represent the power and phase profiles along with the static spin configurations at the various bias fields.

## 2. Results and Discussion

The FESEM images of the nanodot arrays (Fig. 1(a) and (b)) vividly illustrate the distinctive rounded corners and shape asymmetry of the individual nanodots. It is important to note that the dimensions and separations of these individual elements exhibit variations within a range of ±10 nm. In Fig. 1(c), we present typical raw FMR spectra from the two different samples, highlighting discernible peaks corresponding to SW modes. To establish the material parameters according to the Kittel formula[50], we initially measured the bias field-dependent FMR spectrum of a 20-nm-thick continuous Py film which is presented in Fig. 1(d) along with the theoretical fit with the Kittel formula as given by:

$$f = \frac{\mu_0 \gamma}{2\pi} \sqrt{(H + H_k)(H + H_k + 4\pi M_s)} \qquad [1]$$

Here, the uniform precession frequency is denoted by (*f*), vacuum magnetic permeability is denoted by ($\mu_0$), while the other symbols carry the same meaning as described before in this article. The extracted material parameters are as follows: $\gamma$ = 17.95 MHz/Oe, $M_s$ = 800 emu/cm³, and $H_k$ = 0. Subsequently, we explore the bias-field-dependent FMR spectra for both diamond and triangular-shaped nanodot arrays within the range of 0° ≤ φ ≤ 30°, as presented in Figs. 2(a) to 2(f) and Figs. 3(a) to 3(d), respectively.

Several noteworthy features emerge from our analysis of the observed SW spectra. The diamond nanodot array manifests three distinct SW modes, with the lowest-frequency mode exhibiting greater prominence in comparison to the higher-frequency modes. In contrast, the triangular nanodot array demonstrates two SW modes, with the lower-frequency mode displaying greater intensity. An observable but small asymmetry is evident in the SW modes, particularly between positive and negative bias field values even at φ = 0°, attributed to structural imperfections in the both the arrays of nanodots. Notably, in the diamond nanodot array at φ = 5° when transitioning from positive to negative bias fields, the SW spectra exhibit right-skewed characteristics. Conversely, during the transition from negative to positive bias fields, they display a left-skewed profile. We observe sudden magnetic switching events, leading to a swift transition of mode frequencies to higher values at φ = 15° and 30°. Intriguingly, the diamond-shaped nanodot structures exhibit mode softening phenomenon at lower bias fields, primarily pronounced at lower values of φ and gradually diminishing as φ increases. In contrast, the triangular-dot array does not exhibit such mode softening phenomenon, but exhibits mode anti-crossing at φ = 15° and 30°. The anti-crossing field and strength shows change with the variation of bias field angle.

This comprehensive characterization provides valuable insights into the magnetization dynamics and SW behaviour within non-ellipsoidal 2D nanodot arrays, setting the stage for a deeper exploration of the results and their implications. Continuing our exploration of the anisotropic behaviour within the nanodot arrays, we present surface plots at various φ for both the diamond and triangular-shaped arrays, as shown in Fig. 2 and Fig. 3, respectively.

*Diamond Nanodot Array:* At φ = 0°, we observe the presence of mode softening phenomenon occurring between bias fields of 250 Oe and -200 Oe. The magnetic switching of the lowest and the highest frequency modes (Mode 1 and Mode 3, respectively) exhibits a gradual nature. Conversely, the magnetic switching of the intermediate mode (Mode 2) displays a sudden jump in frequency. When we adjust φ to 5°, the mode softening phenomenon slightly diminishes, and the switching field shifts to -250 Oe, with magnetic switching in all the modes becoming more coherent. Remarkably, at φ = 15°, the mode softening almost vanishes, and we observe a shift in the switching field to 300 Oe, accompanied by coherent mode switching. Transitioning to φ = 30°, we find that no qualitative changes occur in comparison to φ = 15°. Intriguingly, at φ = 5°, when we measure the SW spectrum from negative to positive bias field, it resembles a mirror image of the spectrum observed at φ = 5° when measured from positive to negative field. Similarly, the SW spectrum at φ = 15°, when measured from negative to positive bias field, mirrors the spectrum observed when measured from positive to negative field at the same angle. This phenomenon displays remarkable reconfigurability of the SW spectra with subtle variation of bias field configuration.

*Triangular Nanodot Array:* At φ = 0°, we do not observe mode softening phenomenon, and magnetic switching occurs at -200 Oe with coherency. At φ = 15°, we notice no significant qualitative differences from φ = 0°, apart from the appearance of a mode anti-crossing at 1400 Oe and -1400 Oe, which leads to mode hybridisation. Additionally, the switching field shifts to -100 Oe. At φ = 30°, the anti-crossing behaviour becomes more pronounced between the two lower-frequency modes (Mode 1 and Mode 2) and it undergoes a downshift in the magnetic field to 800 Oe and -600 Oe, respectively. Moreover, the anti-crossing gap ($2g$) increases from 1.4 GHz in both +H and -H to 2.47 GHz in +H and 1.73 GHz in -H as φ changes from 15° to 30°. The coupling strength of modes is defined by:

$$g = g_0\sqrt{N} \qquad [2]$$

Here, $g$ represents the overall coupling strength, $g_0$ is the coupling strength between individual spins and $N$ is the number of spins. Interestingly, a fourth mode emerges in the higher frequency range, a feature that was not observed at φ = 0° and φ = 15°. At φ = 15°, when we saturate from negative to positive bias field, the SW spectrum mirrors that when saturating from positive to negative field, similar to the behaviour of the diamond nanodot array in Fig. 2(c). The SW mode anti-crossing indicates the presence of magnon-magnon coupling in triangular nanodots, which is absent in diamond nanodots most likely due to the difference in internal fields for two different shapes. These observations underscore the profound influence of shape anisotropy and asymmetry in contributing to the various phenomena discerned in the experimental data. These findings provide essential insights into the complex interplay between shape and bias field, shedding light on the intricate behaviour of SW modes within these nanodot arrays.

To gain deeper insights into the experimental results, we conducted micromagnetic simulations using mumax[3] software, which is widely recognized for its effectiveness in modelling magnetic phenomena. For dynamic simulations, a Gilbert damping constant ($\alpha$) of 0.008 was employed, while for static calculations, $\alpha$ was set to 0.99 for fast relaxation of the magnetization to its equilibrium state. The detailed methodologies underpinning these simulations are extensively documented elsewhere for reference[22]. In Figs. 2(a) to 2(f) and Fig. 3(a) to 3(d), it is apparent that the micromagnetic simulations aptly replicate the experimental results, offering a qualitative alignment. Crucially, the essential features captured in the simulations are denoted by the dotted symbols, closely mirroring the experimental observations. Expanding on our investigation, we delved into the static magnetization properties, showcased in Figs. 2(a) to 2(b) and Fig. 3(a) to 3(f) through hysteresis loop calculations. Notably, these hysteresis loops are integrated with the SW spectra presented in the aforementioned figures. A noteworthy observation is the nearly concordant alignment of the switching fields (S.F.) obtained from these hysteresis calculations with the SW mode switching fields discerned in the experimental and dynamic studies. Tables 1(a) and 1(b) provide a comprehensive compilation of the switching fields derived from both the SW spectra and hysteresis calculations, further underlining the robust correlation between the static and dynamic aspects of our investigation. This comprehensive approach, combining experimental data, dynamic simulations, and static magnetization properties, yields a coherent and well-supported assessment of the intricate SW behaviour within the non-ellipsoidal nanodot arrays.

**Table 1(a).** Magnetic Switching Fields for Diamond Nanodot Array at Bias Field Angle $0° \leq \varphi \leq 30°$

| Angle (φ) | 0° | 5° | 15° | 30° | 5° -ve to +ve saturation | 15° -ve to +ve saturation |
|---|---|---|---|---|---|---|
| SW Spectra S.F. (Oe) | -200 | -250 | -300 | -200 | 350 | 400 |
| Hysteresis S.F. (Oe) | -215 | -368 | -449 | -129 | 373 | 449 |

**Table 2(b).** Magnetic Switching Fields for Triangular Nanodot Array at Bias Field Angle $0° \leq \varphi \leq 30°$

| Angle (φ) | 0° | 15° | 30° | 15° -ve to +ve saturation |
|---|---|---|---|---|
| SW Spectra S.F. (Oe) | -200 | -100 | -100 | 150 |
| Hysteresis S.F. (Oe) | -249 | -241 | -86 | 246 |

Referring to the hysteresis loops observed in the diamond nanodot array: At $\varphi = 0°$, the magnetic switching manifests as a gradual transition. However, at $\varphi = 5°$, 15°, and 30°, the magnetic switching appears more coherent. Notably, at $\varphi = 5°$ and 15°, when the bias field is applied from negative to positive saturation, the switching also appears to be nearly coherent. In the case of the triangular nanodot array: at $\varphi = 0°$ and 15°, the magnetic switching is near coherent. Likewise, when $\varphi = 15°$ and the bias field is applied from negative to positive saturation, the switching behaviour remains coherent. However, at $\varphi = 30°$, the switching is observed to be gradual, with an entanglement in the loop near the saturation regions.

      To gain a deeper understanding of these magnetic switching behaviours, we conducted simulations of the static magnetic configurations at various magnetic field values for $\varphi = 0°$ and $\varphi = 15°$, as depicted in Figs. 4, 5, 6, and 7. For the diamond nanodot structure: At $\varphi = 0°$, we observe that the spins are strongly oriented towards the bias field angle at H = 1000 Oe, exhibiting an Onion-like spin texture due to the shape geometry of the diamond nanodots, and the bias field influence being perpendicular to the specific corners of the diamond nanodots. As we tune the bias field and gradually reduce it, the uniformity of spin configuration weakens. However, Kittel-like behaviour in SW is retained from H = 1000 Oe down to H = 250 Oe. As

the bias field is reduced further and tuned between -200 ≤ H < 250 Oe, it transits into an S-like state due to the weak bias field influence on the spins and the shape symmetry along the field angle. This transition phase is responsible for the deviation from the traditional Kittel-like behaviour and the observed mode softening phenomenon. This is nearly mirrored in the negative field regime with slight asymmetry due to shape deformation giving rise to the W-shaped dispersion of this mode. In contrast, at φ = 15°, when H = 1000 Oe, the structure displays a strongly oriented quasi-S state instead of the previously present Onion-like state. This is due to the bias field angle being more in line with the edges of the diamond nanodot structures. As the bias field is gradually reduced, the uniformity in spin orientations weakens and form an S-like spin texture at H = 200 Oe. This spin texture remains constant till H = -350 Oe, after which, a swift transition occurs and the spins sharply reorient into a reverse-S state. In this case, the interplay of the bias field angle with the nanodot shape geometry does not allow the accommodation of spin configurations apart from the S-type. The presence of a single type of spin configuration leads to the lack of intermediate spin transition phases. This does not allow for the previously observed mode softening phenomenon to occur. However, a sharp transition in the SW mode frequency occurs at H = -350 Oe due to the aforementioned reversal of spin configuration.

For the triangular-shaped array, both at φ = 0° and 15°, only C-like spin configurations are present in forward and reverse states, being strongly oriented at high bias field values. At φ = 0°, the spin textures can be observed to be more symmetric in nature, along the y-axis. This is due to the triangular dot shape symmetry along the y-axis. However, at φ = 15°, while resembling the same C-like states and switching behaviour, the spins are more oriented towards certain edges due to the bias-field influence while becoming slightly asymmetric along the y-axis. Strong domain walls form around the y-axis, nearing reversal. The key reason behind the coherent switching is the presence of a single type of spin arrangement. This is attributed to the absence of inversion symmetry about the x-axis in the triangular nanodots, allowing the elements to accommodate a specific type of spin arrangement at the given azimuthal angles. Conversely, since inversion symmetry is present in the diamond nanodots, the geometry facilitates multiple types of spin configurations, leading to the observed transition phases. The ability to reconfigure these states in the diamond nanodot arrays by varying the bias field angle offers a crucial advantage into the underlying dynamics. The supplementary material displays static magnetic configurations for the remaining bias field angles in Fig. 1S to Fig. 6S. This comprehensive analysis sheds light on the intricate interplay between nanodot geometry, inversion symmetry, and bias field angle, influencing the magnetization switching and static magnetic configurations in the observed nanodot arrays.

In Fig. 4, we present the power and phase profiles for the diamond nanodot array at $\varphi = 0°$. At H = 1000 Oe, Mode 1 exhibits a characteristic similar to Backward Volume (BV; n) modes with quantization numbers n = 3, m = 1 showing monotonic variation of SW frequency with H. This similarity persists till H = 250 Oe. During the transition of the spin configuration within the S-state, Mode 1 displays a mixed BV and Damon-Eshbach (DE; m) characteristic, with quantization numbers evolving to n = 4, m = 2 at H = 50 Oe and subsequently to n = 3, m = 3 at H = -200 Oe before reverting to n = 3, m = 1 at H = -1000 Oe. In Mode 2, we observe a qualitative similarity in the phase transition and mode localization to that of Mode 1. For Mode 3, at strong bias field values, H = 1000 Oe and -1000 Oe, gives rise to a mixed BV and DE characteristic with quantization numbers n = 6, m = 3. Conversely, at H = 250 Oe and 50 Oe, BV behaviour prevails, with quantization numbers n = 6, m = 1. Notably, the quantization numbers at high field values remain similar at corresponding positive and negative fields. Further exploration at $\varphi = 15°$, as depicted in Fig. 5, reveals that Mode 1 assumes n = 4, m = 1 characteristics at H = 1000 Oe and -1000 Oe, displaying mixed BV and DE characteristics at intermediate fields. A similar trend is observed in Modes 2 and 3. Collectively, these analyses unveil that at high bias field values, the SW mode characteristics predominantly exhibit BV-like behaviour. Conversely, at lower fields and during transition phases, they manifest mixed characteristics. Importantly, it is evident that the quantization numbers increase with the mode number, underscoring the intricate nature of SW behaviour within the diamond nanodot array.

In Fig. 6, we present the power and phase profiles of the triangular-shaped nanodot array at $\varphi = 0°$. Strikingly, the behaviour of this triangular-shaped array closely mirrors that of its diamond-shaped counterpart. In Mode 1, at H = 1000 Oe and -1000 Oe, BV-like characteristics are evident, displaying quantization numbers n = 4, m = 1, and mixed characteristics emerge in the intermediate fields. This behaviour extends to Modes 2 and 3. Notably, in Mode 3 at H = 0 Oe and H = -300 Oe, the SW characteristics assumes a highly complex nature with quantization numbers n = 7, m = 6, underscoring the intricate dynamics of the triangular-shaped nanodot array. Transitioning to Fig. 7 at $\varphi = 15°$, we observe a deviation from the standalone BV-like characteristics at high bias field. In Mode 1 at H = 1000 Oe and -1000 Oe, the previously BV-like behaviour is replaced by BV-DE mixed characteristics, displaying quantization numbers n = 4, m = 2, enhancing our understanding of the evolution of SW behaviour in response to changes in the bias field and geometry. From the supplementary material Fig 5S, we can see that at φ = 30°, while approaching the anti-crossing field at H = 700 Oe, Mode 1 and Mode 2 display mixed BV-DE characteristics having identical quantization numbers n = 3, m = 2, resembling a similar phase. This represents the occurrence of a hybridized mode at that field due to the magnon-magnon coupling. For additional power and phase profiles, we refer readers to the supplementary material, providing a more comprehensive view of the intricacies of SW behaviour in the triangular-shaped nanodot array.

## 3. Conclusion

In summary, we have investigated bias-field dependent magnetization dynamics and static magnetic configurations of $Ni_{80}Fe_{20}$ non-ellipsoidal 2D MCs of diamond and triangular-dot shape at various bias-field angles using broadband ferromagnetic resonance spectroscopy. The SW modes show interesting variation in the diamond nanodot arrays at different bias-field angles. At φ = 0°, we noted a mode softening phenomenon occurring at lower bias-field magnitudes, which gets wiped out at φ = 15°. This tunability is not present in the triangular counterpart. However, here a SW mode anti-crossing is observed at φ = 15°, which undergoes a downshift in field with an increase in the anti-crossing gap at φ = 30°. In the static magnetic configurations, the diamond nanodot array undergoes a transition phase from onion-state to S-state as we decrease the magnitude of the bias field. The triangular nanodot array does not accommodate the transition phases, leading to no mode softening and less tunability. The spatial power and phase maps of the SW modes display mixed BV-DE modes as well as BV-modes. Mode softening and magnetic switching occur where BV-DE like mixed characteristics is present with different quantization numbers. With change in dot shape and bias field angle, qualitative changes in the SW spectra are observed. Calculated hysteresis loops indicate the magnetic switching fields as a static property of the nanodot structures. The large reconfigurability of SW modes in the diamond and triangular nanodot arrays with bias field and angle offer new possibilities for the advancement of magnetic memory, storage, logic and hold significance in designing propagating spin wave based future magnonic devices.

**Acknowledgements**

SB acknowledges Technical Research Centre (TRC), S. N. Bose National Centre for Basic Sciences for funding and research fellowship. We thank Dr. Yoshichika Otani for providing the samples.

**Conflict of Interest**

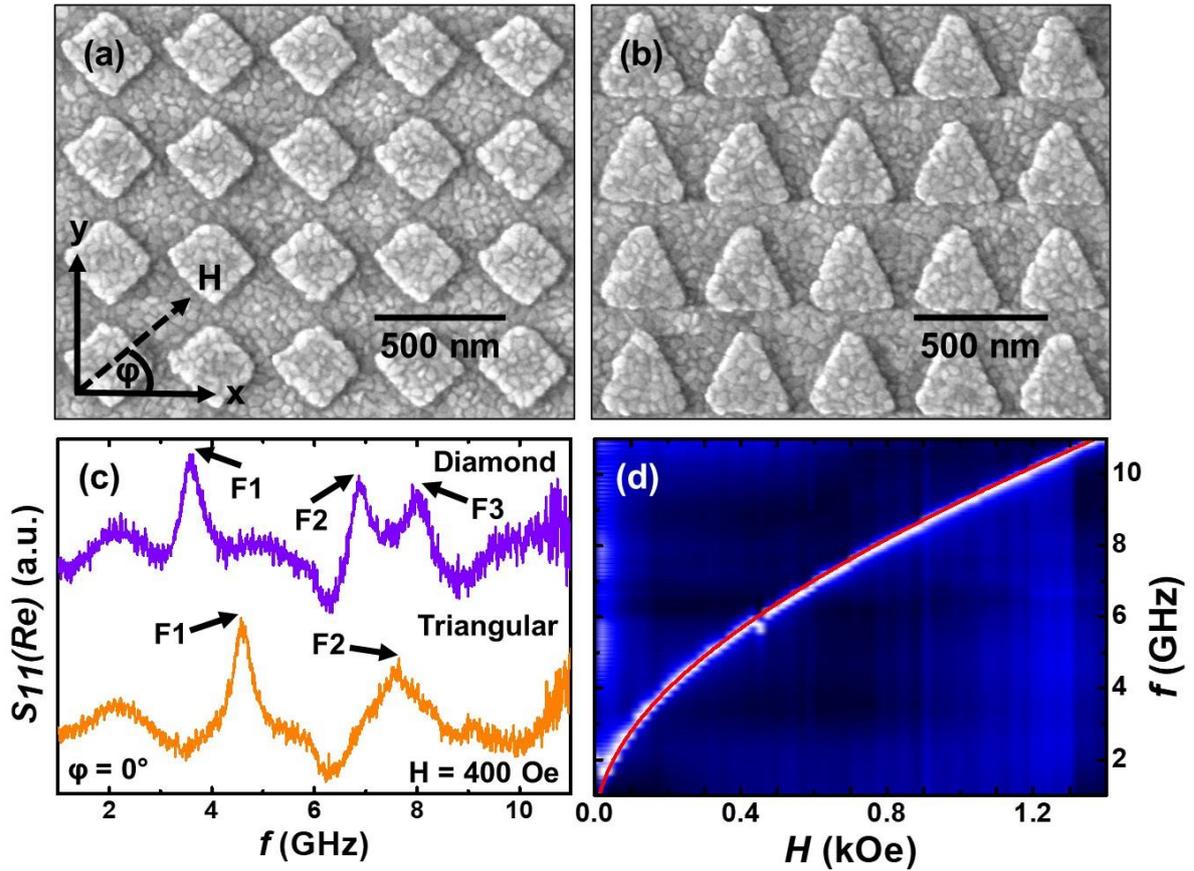

**Fig. 1.** FESEM images of Py nanodot arrays of dot dimensions 300 × 330 nm spaced at 75 nm, of (a) diamond and (b) triangular dot shape. Fig. (c) represents the Real part of $S_{11}$ parameter of as a function of frequency, with arrows representing the SW modes, and (d) represents the Kittel-behaviour of Py thin film, where the surface plot is the experimental result, while the red solid line represents the theoretical fit.

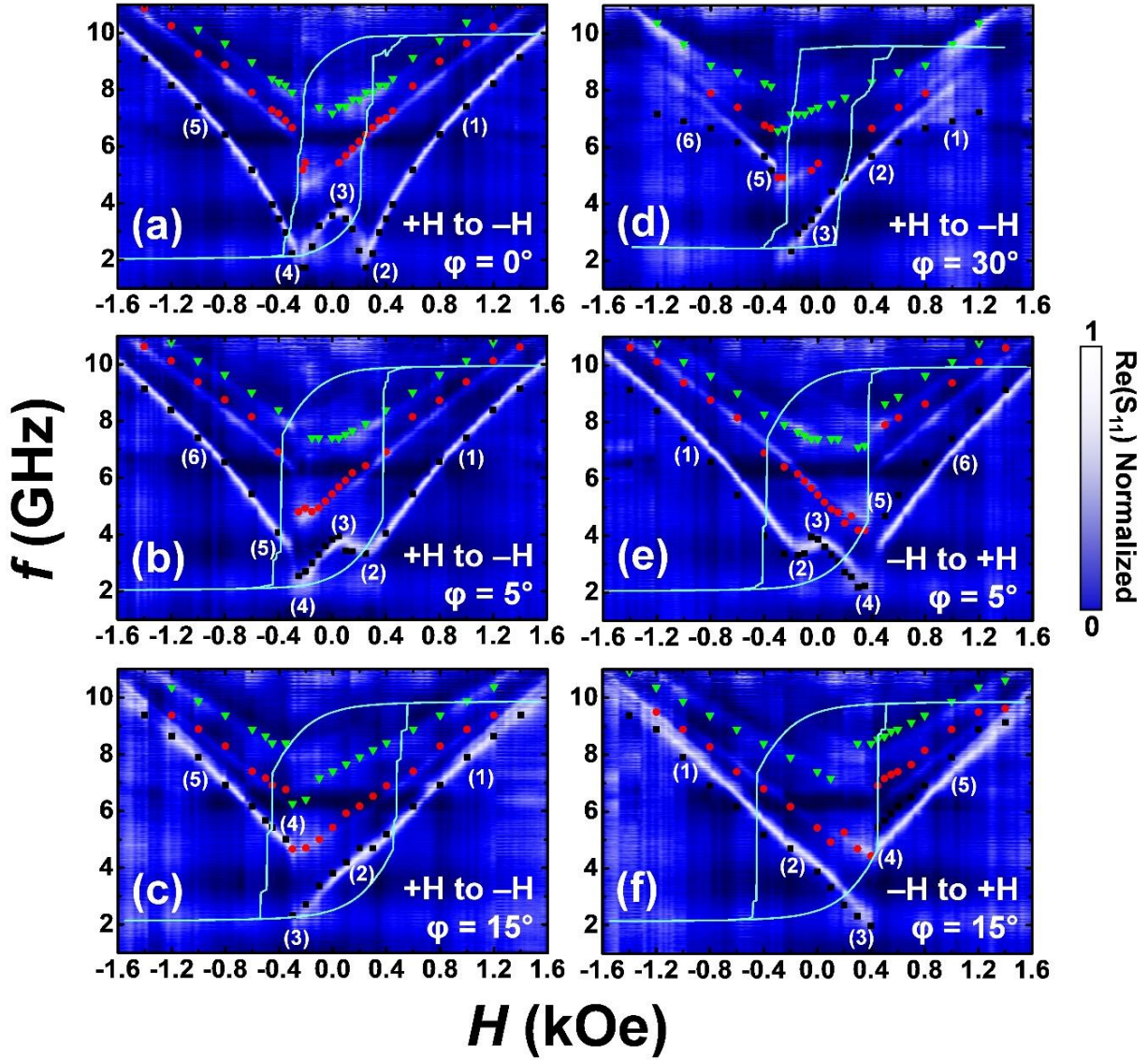

**Fig. 2.** Surface plots of bias-field dependent SW dispersion spectra of diamond nanodot array at φ = (a) 0°, (b) 5°, (c) 15°, (d) 30° when bias-field is saturated from +ve to -ve and φ = (e) 5°, (f) 15° when bias-field is saturated from -ve to +ve respectively, while power (P) = -15 dBm. The simulated SW modes are represented by symbols and simulated hysteresis loops of each configuration are shown on top of the surface plots.

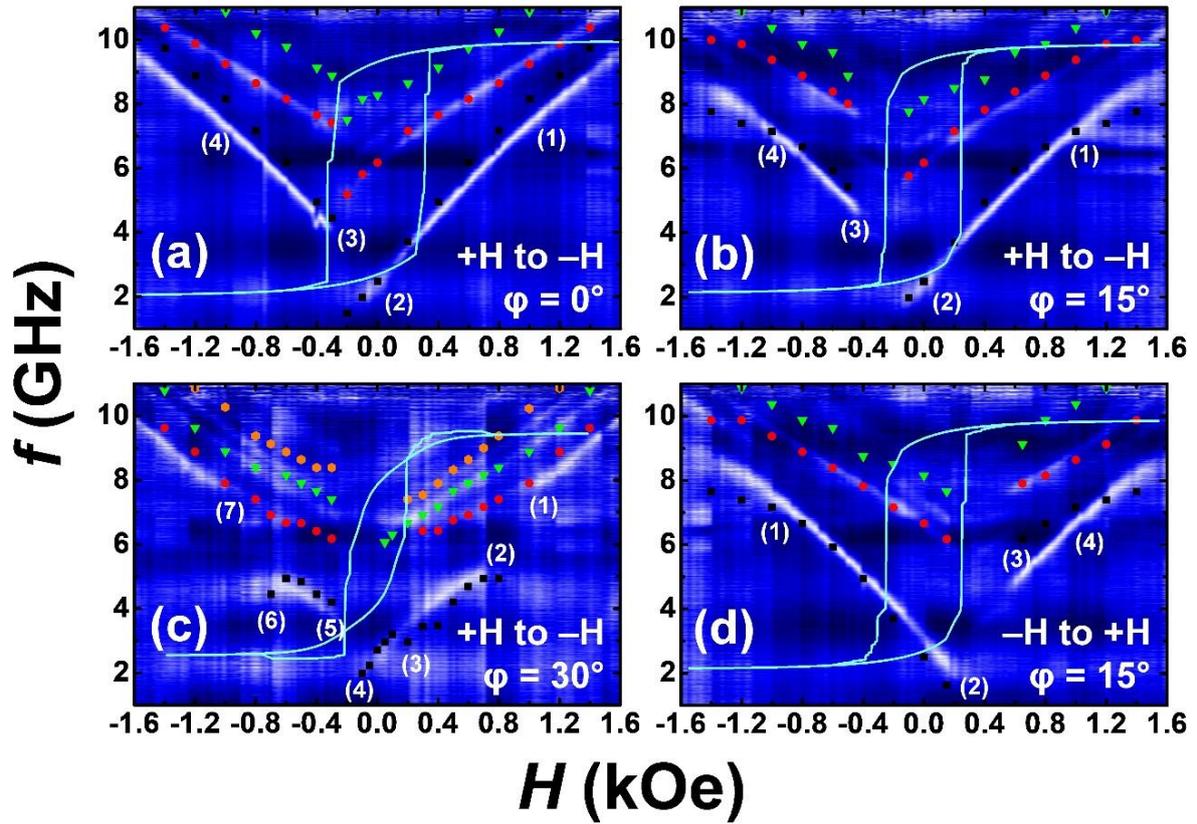

**Fig. 3.** Surface plots of bias-field dependent SW dispersion spectra of triangular nanodot array at φ = (a) 0°, (b) 15°, (c) 30° when bias-field is saturated from +ve to -ve and φ = (d) 15° when bias-field is saturated from -ve to +ve respectively, while power (P) = -15 dBm. The simulated SW modes are represented by symbols and simulated hysteresis loops of each configuration are shown on top of the surface plots.

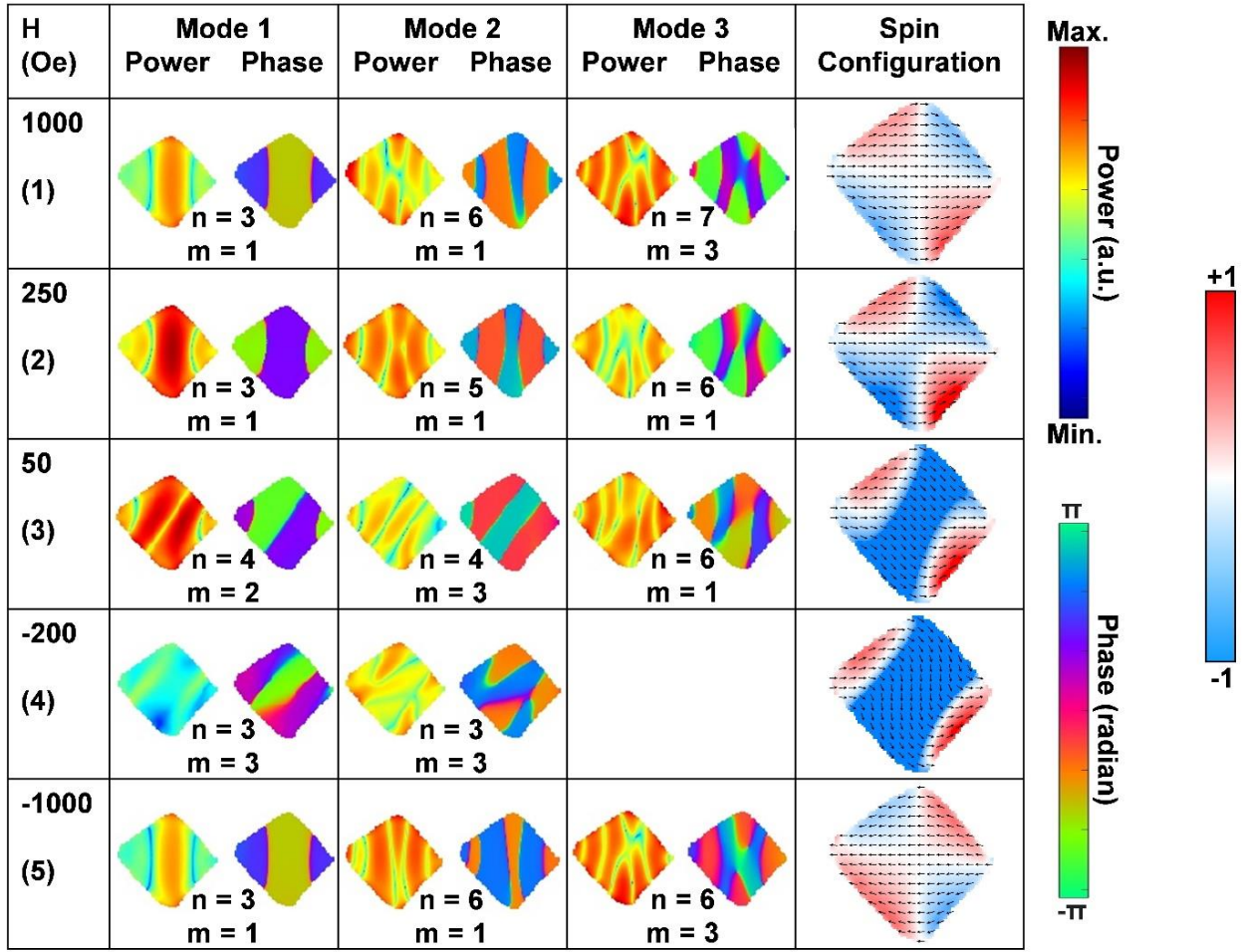

**Fig. 4.** Simulated spatial map of power and phase of the SW modes of diamond nanodot array at bias-field φ = 0° along with static spin configurations at various field values. The colormaps for SW mode profiles and spin configurations are shown at the right side.

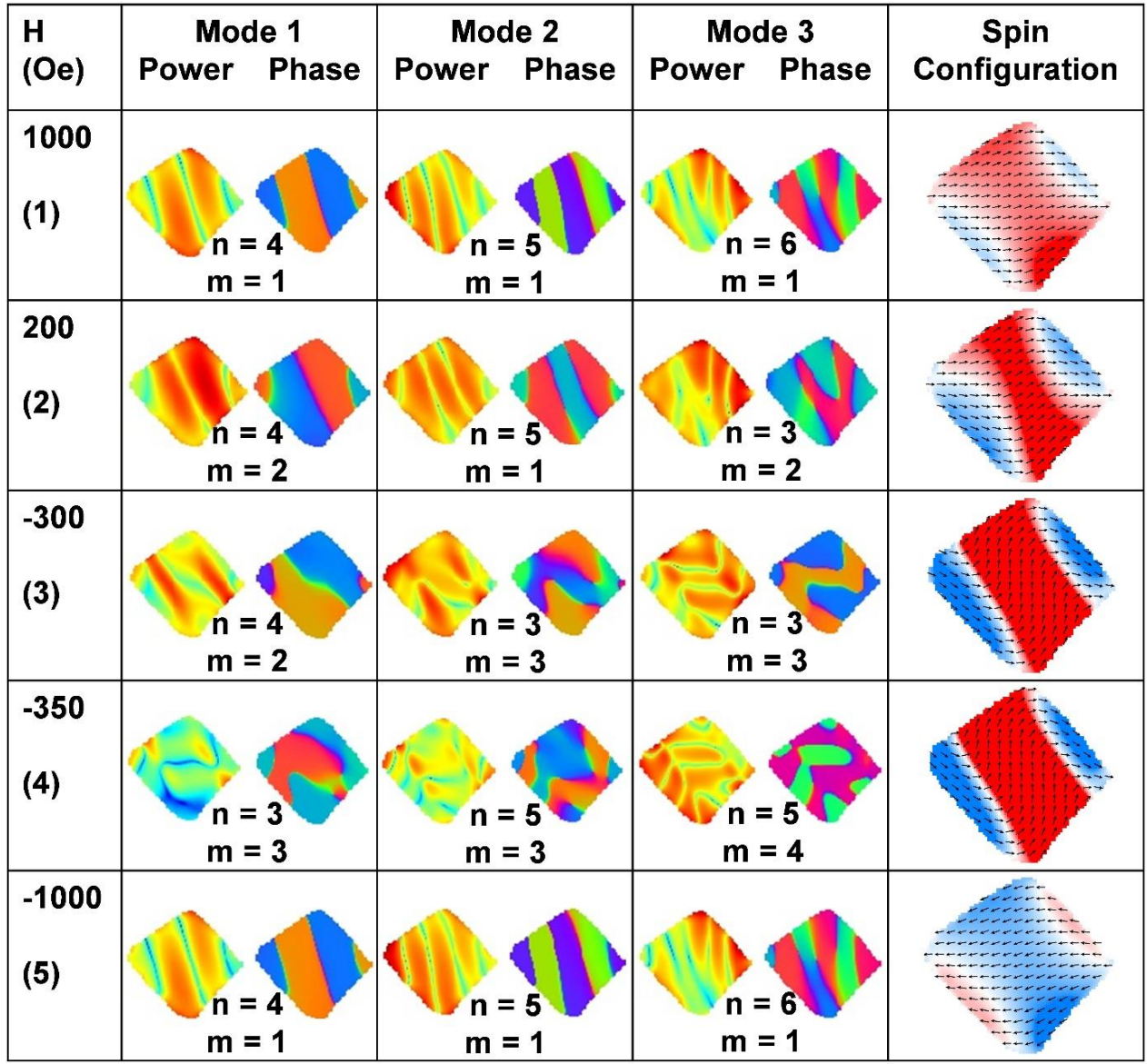

**Fig. 5.** Simulated spatial map of power and phase of the SW modes of diamond nanodot array at bias-field φ = 15° along with static spin configurations at various field values. The colormaps for SW mode profiles and spin configurations are shown in Fig. 4.

| H (Oe) | Mode 1 Power  Phase | Mode 2 Power  Phase | Mode 3 Power  Phase | Spin Configuration |
|---|---|---|---|---|
| 1000 (1) | 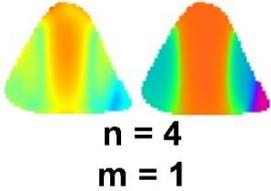 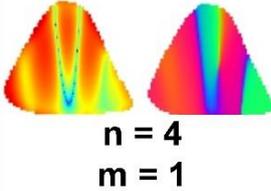 n = 4 m = 1 | 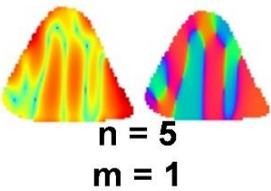 n = 4 m = 1 | 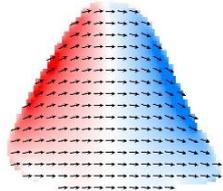 n = 5 m = 1 | 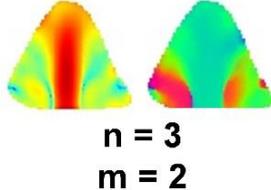 |
| 0 (2) | 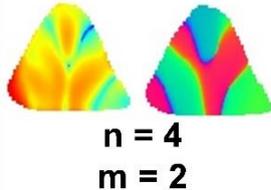 n = 3 m = 2 | 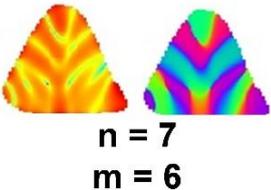 n = 4 m = 2 | 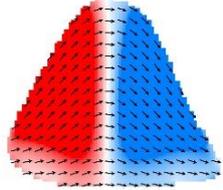 n = 7 m = 6 | 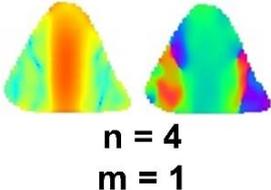 |
| -300 (3) | 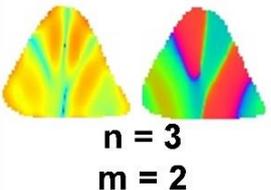 n = 4 m = 1 | 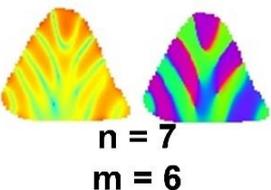 n = 3 m = 2 | 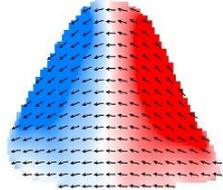 n = 7 m = 6 | 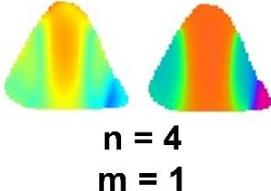 |
| -1000 (4) | 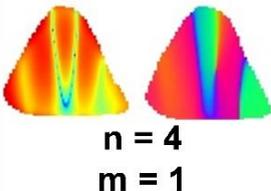 n = 4 m = 1 | 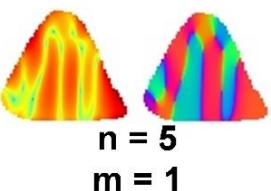 n = 4 m = 1 | 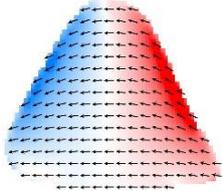 n = 5 m = 1 |  |

**Fig. 6.** Simulated spatial map of power and phase of the SW modes of triangular nanodot array at bias-field φ = 0° along with static spin configurations at various field values. The colormaps for SW mode profiles and spin configurations are shown in Fig. 4.

| H (Oe) | Mode 1 Power Phase | Mode 2 Power Phase | Mode 3 Power Phase | Spin Configuration |
|---|---|---|---|---|
| 1000 (1) | 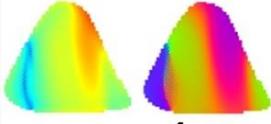 n = 4, m = 2 | 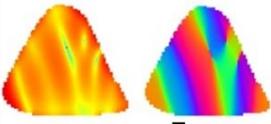 n = 5, m = 4 | 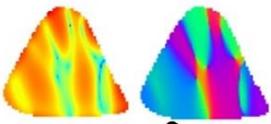 n = 3, m = 2 | 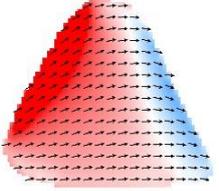 |
| 0 (2) | 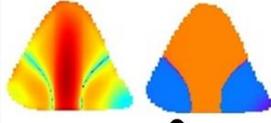 n = 3, m = 2 | 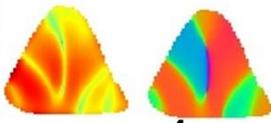 n = 4, m = 3 | 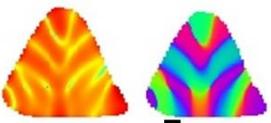 n = 7, m = 6 | 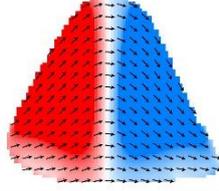 |
| -500 (3) | 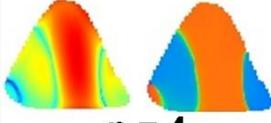 n = 4, m = 2 | 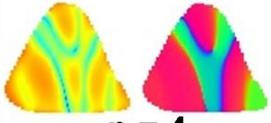 n = 4, m = 4 | 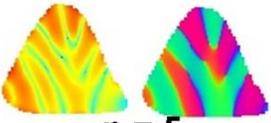 n = 5, m = 4 | 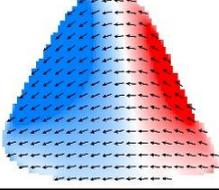 |
| -1000 (4) | 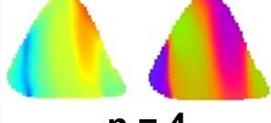 n = 4, m = 2 | 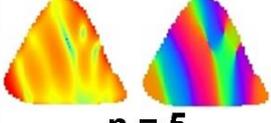 n = 5, m = 4 | 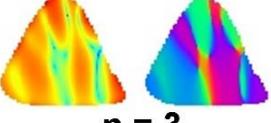 n = 3, m = 2 | 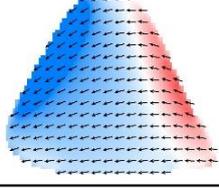 |

**Fig. 7.** Simulated spatial map of power and phase of the SW modes of triangular nanodot array at bias-field $\varphi = 15°$ along with static spin configurations at various field values. The colormaps for SW mode profiles and spin configurations are shown in Fig. 4.